\documentclass[12pt,superscriptaddress,nofootinbib,onecolumn,aps,prd]{revtex4-1}

\usepackage[cp1251]{inputenc}
\usepackage[russian]{babel}

\usepackage{amsmath}
\usepackage{mathtools}

\def\ga{\mathrel{\mathpalette\fun >}}
\def\fun#1#2{\lower3.6pt\vbox{\baselineskip0pt\lineskip.9pt
\ialign{$\mathsurround=0pt#1\hfil##\hfil$\crcr#2\crcr\sim\crcr}}}

\begin{document}
\title{Некоторые результаты, полученные теоретиками
ИТЭФ за семьдесят лет работы Института}
\author{М.И. Высоцкий}
\email{vysotsky@itep.ru}
\affiliation{Институт теоретической и экспериментальной физики  им. А.И.Алиханова}
\affiliation{Московский физико-технический институт}
\affiliation{Московский инженерно-физический институт}
\author{А.Д. Долгов}
\email{dolgov@fe.infn.it}
\affiliation{Университет Феррары (Италия)} 
\affiliation{Новосибирский государственный университет}
\author{В.А. Новиков}
\email{novikov@itep.ru}
\affiliation{Институт теоретической и экспериментальной физики  им. А.И.Алиханова}
\affiliation{Московский физико-технический институт}
\affiliation{Московский инженерно-физический институт}

\maketitle

\section*{Введение}

Перед организованной в декабре 1945 года Лабораторией №3
(впоследствии переименованной в Институт теоретической и
экспериментальной физики, сейчас -- ИТЭФ им. А.И.~Алиханова) были
поставлены конкретные задачи по созданию ядерных реакторов.
Позднее, в конце 50-х -- 60-х годов перед Институтом стояла задача
проектирования протонных ускорителей с жесткой фокусировкой.
Теоретики ИТЭФ внесли важный вклад в решение этих задач, однако
прикладная тематика в работе теоротдела всегда сопровождалась
фундаментальными исследованиями. Полученным в теоротделе
фундаментальным результатам и посвящена эта статья.

В декабре 1945 года руководителем всех теоретических работ
Лаборатории №3 был назначен Лев Давидович Ландау (1908--1968,
академик АН СССР), а в 1946 году главой теоротдела стал его ученик
Исаак Яковлевич Померанчук (1913--1966, академик АН СССР).
Л.Д.~Ландау до 1958 года работал в ИТЭФ по совместительству и
регулярно участвовал в семинарах. В штат теоротдела с момента
организации ИТЭФ наряду с Померанчуком входили Владимир Борисович
Берестецкий и Алексей Дмитриевич Галанин. 

В конце этого краткого
введения дадим ссылку на статью \cite{Okun}, посвященную
выполненным в ИТЭФ теоретическим работам и написанную в связи с
шестидесятилетием Института.

\section*{Квантовая электродинамика}

В 1939--1946 гг. И.Я.~Померанчуком разработана теория излучения
релятивистских электронов в магнитном поле (магнитотормозное, или
синхротронное, излучение) \cite{1,2}. В применении к космическим
лучам это излучение устанавливает верхний предел для энергий, с
которыми могут поступать на поверхность земли электроны и
позитроны, входящие в состав первичных космических лучей. Это же
излучение делает невозможным построение кольцевых $e^+
e^-$-коллайдеров на очень высокие энергии: планируемые в настоящее
время $e^+ e^-$-коллайдеры на полную энергию в несколько сотен ГэВ
(несколько ТэВ) являются либо линейными (ILC, международный
линейный коллайдер; CLIC, компактный линейный коллайдер), либо
имеют очень большой радиус при относительно низкой энергии
(FCC-$ee$, будущий кольцевой коллайдер с длиной кольца 100
километров). В настоящее время электронные кольцевые накопители
используются как источники синхротронного излучения, эксперименты
с которым дают существенный вклад в развитие атомной и
молекулярной физики, физики твердого тела, в изучение катализа, в
материаловедение и биофизику.

Для нахождения волновых функций фотона в 1947 году
В.Б.~Берестецким была разработана теория шаровых векторов и с ее
помощью построена теория бета-гамма корреляций при распадах ядер
\cite{3}.

В 1948 году Померанчук обратил внимание на то, что два сорта
позитрониев ($e^+ e^-$-атомов) -- ортопозитроний, в котором спины
электрона и позитрона складываются в суммарный спин единицу, и
парапозитроний, в котором они складываются в суммарный спин ноль,
-- должны иметь существенно разные времена жизни (речь идет об
основных состояниях позитрониев с нулевым орбитальным моментом)
\cite{4}. Дело в том, что для ортопозитрония двухфотонная
аннигиляция запрещена, и он распадается на три фотона, живя
примерно в тысячу раз дольше, чем парапозитроний. Невозможность
распада ортопозитрония на два фотона легко пояснить, используя
сохранение зарядовой четности ($C$) в электромагнитных
взаимодействиях. $C$-четность $e^+ e^-$-пары равна $(-1)^{l+s}$,
где $l$ -- орбитальный момент пары, а $s$ -- суммарный спин. В
основном состоянии $l=0$, поэтому у парапозитрония оно $C$-четно,
а у ортопозитрония -- $C$-нечетно. Отрицательная зарядовая
четность фотона запрещает распад $C$-нечетного основного состояния
ортопозитрония на два фотона. Аналогичный механизм приводит к
большому времени жизни $J/\psi$-мезона, являющегося связанным
состоянием очарованных $c\bar c$-кварков. Имеющий спин единицу
$J/\psi$ не может распадаться на два глюона; распад идет на три
глюона, и его вероятность подавлена кубом аналогичной постоянной
тонкой структуры $\alpha$ константы сильного взаимодействия
$\alpha_s$. Тот же механизм объясняет узость $\Upsilon$-мезона,
состоящего из прелестных $b\bar b$-кварков (важна также большая
масса $c$- и $b$-кварков; ``константа'' $\alpha_s$ уменьшается с
ростом характерной энергии, роль которой играет масса тяжелого
кварка).

Ознакомившись с результатом Померанчука об отсутствии распадов
ортопозитрония на два фотона, Ландау в том же 1948 году доказал
общую теорему, согласно которой два фотона не могут находиться в
состоянии с полным моментом, равным единице \cite{5}. В литературе
это утверждение носит название теоремы Ландау--Янга; Ч.Н.~Янг
(США) пришел к такому же утверждению в 1950 году \cite{6}. Эта
теорема сыграла важную роль в определении квантовых чисел бозона
$H$ с массой 125 ГэВ, открытого в 2012 году на LHC: детектирование
распада $H$ на два фотона доказало, что его спин не может
равняться единице (наиболее вероятно, согласно экспериментальным
данным, значение спина ноль, как и должно быть для бозона Хиггса).

В работе Берестецкого и Ландау 1949 года \cite{7} получен
гамильтониан, описывающий систему $e^+ e^-$ с точностью до членов
$\sim v^2/c^2$. В работе Берестецкого того же года этот
гамильтониан использован для определения тонкой структуры уровней
позитрония, в частности, найдено, что основной уровень
ортопозитрония лежит выше основного уровня парапозитрония на
величину $\Delta = (4/3+1)m\alpha^4/4$ (второй член в скобке
соответствует аннигиляционной диаграмме) \cite{8}. В этой же
работе отмечена особенность эффекта Зеемана в позитронии: линейный
по магнитному полю сдвиг уровней отсутствует; в магнитном поле
происходит смешивание орто- и парапозитрониев. Экспериментальное
исследование влияния магнитного поля на распад позитрония
позволило в начале 50-х годов измерить  орто-пара-расщепление
\cite{9}. Теоретическая точность вычисления $\Delta$ находится
сейчас на уровне поправок $\sim \alpha^7$, отвечающих трехпетлевым
диаграммам, и согласуется с экспериментальными результатами,
имеющими сравнимую точность \cite{10}.

В 1951 году Берестецким установлена фундаментальная теорема о
противоположности внутренней пространственной четности фермиона и
антифермиона \cite{11}. Наряду с позитронием она важна также и при
определении пространственной четности мезонов, являющихся
связанными состояниями пары кварк-антикварк: $s$-волновые
состояния с полным спином 0 и 1 (скажем, $\pi$- и $\rho$-мезоны)
$P$-нечетны, так как $(-1)^{l+1} = -1$.

В 1952 году Галанин и Померанчук рассмотрели лэмбовский сдвиг в
мюонном водороде -- атоме, в котором место электрона занимает мюон
\cite{12}. Вследствие своей большой массы мюон находится в
$m_\mu/m_e \approx$ 210 раз ближе к ядру, чем электрон. Это
приводит к качественному эффекту -- в отличие от обычного водорода
в лэмбовском сдвиге мюонного водорода доминирует изменение
кулоновского потенциала протона на малых расстояниях, вызванное
ростом постоянной тонкой структуры $\alpha$. К этому росту
потенциала более чувствительны $s$-уровни, поэтому уровень $2s$
оказывается связанным сильнее, чем $2p$, в то время как в обычном
водороде уровень $2s$ за счет лэмбовского сдвига поднимается вверх
относительно $2p$-уровня. К сдвигу атомных уровней приводит также
конечный зарядовый радиус протона $r_p$. Извлекаемое из спектра
мюонного водорода с учетом лэмбовского сдвига значение $r_p$
сильно (на уровне 5-8 стандартных отклонений) отличается от
значения, следующего из спектра обычного водорода и экспериментов
по $ep$-рассеянию. В этом противоречии заключается современная
проблема зарядового радиуса протона \cite{13}.

Сечение процесса аннигиляции $e^+ e^- \to \mu^+ \mu^-$,
вычисленное Берестецким и Померанчуком в 1954 году, используется
для нормировки сечений, измеряемых на $e^+ e^-$-коллайдерах
\cite{14}.

В работе 1956 года В.В.~Судаков выделил и вычислил так называемые
дважды логарифмические члены, определяющие асимптотическое
поведение вершинных диаграмм в квантовой электродинамике при
высокой энергии в произвольном порядке теории возмущений и
просуммировал эти члены \cite{15}. Появление дважды логарифмов
связано с тем, что переносчиком взаимодействия в КЭД является
фотон, частица со спином один. Согласно Стандартной Модели,
переносчиками сильных и слабых взаимодействий также являются
частицы со спином один, глюоны и $W^\pm$- и $Z$-бозоны,
соответственно. Поэтому формфактор Судакова играет важную роль
наряду с квантовой электродинамикой также в сильных и слабых
взаимодействиях при высоких энергиях. В качестве относительно
недавнего применения квантовоэлектродинамического формфактора
Судакова отметим, что учет виртуальных и реальных фотонов ведет к
подавлению инклюзивного сечения рождения $Z$-бозона в $e^+
e^-$-аннигиляции фактором ${\rm
exp}\left[-\frac{2\alpha}{\pi}\ln\left(\frac{M_Z^2}{m_e^2}\right)
\ln\left(\frac{M_Z}{\Gamma_Z}\right)\right] \approx 0.7$ --
малость константы $\alpha$ компенсируется дважды логарифмическим
множителем благодаря большой массе $Z$-бозона, $M_Z \approx 91$
ГэВ.

Согласно решению уравнения Дирака для электрона в поле точечного
ядра с зарядом $Z$ энергия основного состояния водородоподобного
иона, равняясь $W=m_e\sqrt{1-(\alpha Z)^2}$, обращается в ноль при
$Z=137$ и при б\'{о}льших $Z$ становится чисто мнимой. В работе
1945 года Померанчук с Я.А.~Смородинским отметили, что учет
конечного размера ядра устраняет корневую особенность в
зависимости энергии от $Z$ \cite{16}. С ростом $Z$ энергия
становится отрицательной и при некотором заряде, названном в
статье критическим, достигает величины $-m_e$. При этом, как
отметили С.С.~Герштейн и Я.Б.~Зельдович \cite{17}, энергетически
возможным становится рождение двух $e^+ e^-$-пар из вакуума, при
котором электроны занимают атомный уровень с энергией $-m_e$, а
позитроны уходят на бесконечность. В работах В.С.~Попова \cite{18}
проведено вычисление критического заряда, уточняющее результат
Померанчука и Смородинского; показано, что волновая функция сильно
связанного электрона сосредоточена вблизи ядра на расстояниях
$\sim 1/m_e$, найдено время, за которое $e^+ e^-$-пары рождаются
из вакуума. Эти и многие другие результаты изложены в обзоре
Зельдовича и Попова \cite{19}. В дальнейших работах В.С.~Попова с
соавторами решается задача о рождении пар при столкновениях двух
тяжелых ядер, когда эффективный заряд адиабатически увеличивается
и достигает критической величины \cite{20}. Так как $Z_{\rm
cr}\approx 175$, то обычно обсуждается столкновение двух ядер
урана.

Величина $Z_{\rm cr}$ уменьшается во внешнем магнитном поле, так
как происходит ``поджатие'' электронной орбиты к ядру,
увеличивающее энергию связи электрона. Эффект становится
существенным при полях $B\ga B_0\equiv m_e^2/e$. Согласно
результатам В.Н.~Ораевского, А.И.~Реза и В.Б.~Семикоза (ИЗМИРАН)
\cite{21} при $B\approx 100 B_0$ ядро урана становится
критическим, а при $B= 3\cdot 10^4 B_0$ критическим становится
ядро с $Z=40$. Однако учет поляризации вакуума в сверхсильном
магнитном поле $B > m^2/e^3$, приводящий к экранировке
кулоновского потенциала (В.В.~Усов, А.Е.~Шабад, ФИАН) \cite{22},
качественно меняет описанную картину. Согласно полученным
М.И.~Высоцким, С.И.~Годуновым и Б.~Маше (Франция) результатам
\cite{23}, ядра с $Z < 60$ не становятся критическими ни при каком
$Z$, а ядра с зарядами в интервале $60 < Z < 210$ становятся
критическими при существенно б\'{о}льших магнитных полях, чем без
учета экранировки, и являются критическими в конечном интервале
изменения $B$. Ядра же с $Z>210$ являются критическими при любых
значениях внешнего магнитного поля. Учет конечного размера ядра
важен для определения области критичности.

Основой аналитического описания ионизации атомов, ионов и твердых
тел интенсивным лазерным излучением является работа Л.В.~Келдыша
(ФИАН) \cite{24}. А.М.~Переломовым, В.С.~Поповым и
М.В.~Терентьевым был развит метод мнимого времени для построения
теории многофотонной ионизации атомов интенсивным лазерным светом
\cite{25}. Этот метод широко применялся В.С.~Поповым с соавторами
в физике сильных лазерных полей, в том числе в проблеме рождения
$e^+e^-$-пар лазерным излучением из вакуума \cite{26,261}.

Написанные В.Б.~Берестецким в соавторстве с А.И.~Ахиезером и
Е.М.~Лифшицем и Л.П.~Питаевским монографии ``Квантовая
электродинамика'' \cite{27} и ``Релятивистская квантовая теория''
\cite{28} (переименованная в ``Квантовую электродинамику'' в более
поздних изданиях) на сегодняшний день остаются одними из лучших
учебников в этой области. Первое издание первой книги появилось в
1953 году, и, как написал Ф.~Дайсон (США), ``эта книга является
первой хорошей монографией по квантовой электродинамике и,
вероятно, на долгое время останется лучшей''.

\section*{Ноль заряда и асимптотическая свобода}

Ландау и Померанчук в 1955 году обнаружили, что поляризация
вакуума полностью экранирует конечный точечный заряд в квантовой
электродинамике \cite{29}. Идея об исчезновении заряда была
независимо высказана Е.С.~Фрадкиным (ФИАН). В работах 1955--1956
годов Померанчука \cite{30} и Померанчука, Судакова и
К.А.~Тер-Мартиросяна \cite{31} было установлено, что такое же
поведение заряда имеет место в юкавских теориях. Это явление было
названо ноль заряда (или московский ноль). Малость постоянной
тонкой структуры приводит к тому, что связанные с нулем заряда
проблемы в КЭД начинаются при очень высоких энергиях $\sim
m_e\cdot {\rm exp}(1/\alpha)$ (или на очень малых расстояниях
$\sim {\rm exp}(-1/\alpha)/m_e$). В случае же сильных
взаимодействий большая величина заряда делает квантовую теорию
поля абсолютно неприменимой. Открытие нуля заряда привело к тому,
что в последующие 15 лет развивались методы, основанные на таких
общих принципах, как унитарность и аналитичность матрицы
рассеяния. Теорией поля, основанной на лагранжевом подходе, мало
кто занимался. Впоследствии оказалось, что, как и в случае многих
других ``no-go'' теорем в физике, в утверждении о ноль зарядном
поведении всех квантовых теорий поля имеется исключение. А именно,
в неабелевых калибровочных теориях заряд ведет себя
противоположным образом: с ростом энергии (или уменьшением
расстояния) он падает, имеет место асимптотическая свобода. Это
делает современную теорию сильных взаимодействий, основанную на
неабелевой группе SU(3), самосогласованной на малых расстояниях,
на которых константа $\alpha_s$ становится маленькой. Проблемы
наступают на больших расстояниях, когда заряд становится порядка
единицы и методы теории возмущений не работают. Приходится
ограничиться качественной картиной конфайнмента (пленения) кварков
и глюонов и результатами численных вычислений свойств адронов.

Впервые с противоречащим ноль зарядному поведением заряда в
4-мерной теории столкнулись В.С.~Ваняшин (Днепропетровский
университет) и М.В.~Терентьев в 1965 году при изучении
электродинамики массивных заряженных векторных бозонов
\cite{32}\footnote{Отсутствие зануления заряда в четырехфермионной
теории в двумерном пространстве-времени ранее обнаружил
А.А.~Ансельм (ПИЯФ) \cite{33}.}. В частности, в их работе можно
найти первый коэффициент функции Гелл-Манна -- Лоу КЭД, равный -7
за счет вклада массивных $W^\pm$-бозонов. Амплитуда обнаруженного
в 2012 году на эксперименте распада бозона Хиггса в два фотона
содержит множитель -7+16/9, где -7 - вклад $W^\pm$, а 16/9 --
вклад $t$-кварковой петли. Свой результат по аномальному поведению
заряда авторы \cite{32} связали с неперенормируемостью теории с
массивными векторными бозонами (то, что механизм Хиггса делает
такую теорию перенормируемой, было выяснено значительно позже). В
1968 году И.Б.~Хриплович (ИЯФ, Новосибирск) вычислил бег заряда в
неабелевой калибровочной теории с безмассовыми векторными
частицами, основанной на группе SU(2) \cite{34}. Его ответ для
первого коэффициента функции Гелл-Манна--Лоу оказался равным
-22/3, что превращается в -7 при учете вклада делающей векторные
бозоны массивными голдстоуновской моды -- заряженного бозона
Хиггса: $-22/3 + 1/3 = -7$. Примерно в те же годы в экспериментах
по глубоко неупругому рассеянию было обнаружено, что, если сильные
взаимодействия описываются квантовой теорией поля, то заряд в
такой теории с ростом энергии должен уменьшаться. В начале 70-х
годов основанная на неабелевой группе SU(3) теория с цветными
глюонами и кварками (КХД) уже рассматривалась как теория сильных
взаимодействий. В 1972 году на конференции в Марселе 'т~Хофт в
ходе обсуждений заметил, что согласно его вычислениям в этой
теории заряд с ростом энергии падает. Подробное рассмотрение этих
вопросов, включая анализ эволюции структурных функций протона в
КХД было проделано в работах Политцера \cite{35} и Вильчика и
Гросса (США) \cite{36} 1973 года. С тех пор КХД рассматривается
как правильная теория сильных взаимодействий. В Стандартной Модели
физики элементарных частиц, основанной на группе $SU(3)\times
SU(2)\times U(1)$ имеются как заряды с асимптотически свободным
поведением ($g_3$ и $g_2$), так и с ноль зарядным поведением
($g_1$, а также константы связи бозона Хиггса с кварками и
лептонами). Один из важных принципов построения теорий при высоких
энергиях -- отсутствие в них полюса Ландау, как стало сейчас
называться явление ноля заряда.

\section*{Аномалии}

Иногда симметрии классической теории нарушаются при учете петлевых
поправок. Наиболее известный пример -- аксиальная симметрия,
приводящая к сохранению аксиального тока в квантовой
электродинамике безмассовых электронов, нарушаемая при учете
треугольных диаграмм, описывающих переход аксиального тока в два
фотона через электронную петлю. Рассмотрение дивергенции
нейтрального изотриплетного аксиального тока позволило вычислить
ширину распада $\pi^0 \to 2\gamma$ \cite{38}. В работе
М.В.~Терентьева аналогичное рассмотрение позволило найти амплитуду
перехода $\gamma \to \pi^+ \pi^- \pi^0$ в пределе малых импульсов
пионов \cite{39}. В работе \cite{40} им была предложена
экспериментальная проверка полученного результата в фоторождении
нейтрального пиона при когерентном рассеянии заряженного пиона на
тяжелом ядре, осуществленная на ускорителе ИФВЭ \cite{41}.
Аналогичные процессы с участием $K$-мезонов в пределе малой массы
$s$-кварка были изучены Вессом и Зумино \cite{42}; их описание в
терминах эффективного лагранжиана дано Виттеном \cite{43}.

Причиной аномалий является плохая сходимость интегралов,
отвечающих фейнмановским диаграммам, в области большого импульса
интегрирования. Однако, как отметили А.Д.~Долгов и В.И.~Захаров,
аномальные амплитуды имеют ненулевые мнимые части $\sim
\delta(q^2)$, где $q^2$ -- импульс, входящий в вершину аксиального
тока \cite{44}. Этот инфракрасный аспект аномалии позволяет из их
наличия получить некоторые заключения о спектре адронов, что на
сегодняшний день невозможно сделать на основании лагранжиана КХД,
так как на массовой оболочке взаимодействие является сильным.
Особенность $\sim \delta(q^2)$ в диаграммах с виртуальными
безмассовыми кварками $u$, $d$ и $s$ должны воспроизводиться
диаграммами, в которых по внутренним линиям распространяются
адроны. Единственная возможность реализации этого требования --
иметь в спектре адронов безмассовые в пределе $m_{u,d,s}\to 0$
(псевдо)скалярные состояния. Ими является октет
псевдоголдстоуновских бозонов $\pi^\pm$, $\pi^0$, $K^\pm$, $K^0$,
$\bar K^0$, $\eta$. Это следствие инфракрасного аспекта аномалий
было отмечено Е. Виттеном и С. Коулменом \cite{45}.

\section*{Слабые взаимодействия}

В 1956 году для решения $\theta$ - $\tau$ проблемы в распадах
заряженных каонов Т.Д.~Ли и Янг (США) предположили, что в этих
распадах нарушается P-четность. Для проверки своей гипотезы о
нарушении P-четности в слабых взаимодействиях они предположили,
что в $\beta$-распадах поляризованных ядер может наблюдаться
корреляция импульса рождающихся электронов и спина ядра вида $\bar
s \bar p$. Так как эта корреляция является T-четной, то в силу
CPT-теоремы ее экспериментальное обнаружение означало бы также
нарушение C-четности. Нарушение же C-четности разрушало объяснение
установленной картины распадов нейтральных $K$-мезонов, согласно
которой короткоживущий $K_S$, будучи C-четным, распадался на два
$\pi$-мезона, а C-нечетный $K_L$ жил долго, так как на два
$\pi$-мезона распадаться не мог. Однако, как заметили Б.Л.~Иоффе,
Л.Б.~Окунь и А.П.~Рудик в работе 1957 года \cite{46}, сохранение
T-четности означает сохранение произведения С на P, а этого
достаточно для запрета распадов $K_L \to 2\pi$, так как два
$\pi$-мезона в $s$-волне образуют не только С, но и CP-четное
состояние, $K_L$ же CP нечетен. Таким образом наблюдение
корреляции $\bar s \bar p$ не противоречит существованию
долгоживущего $K_L$, а свидетельствует о нарушении С-четности в
слабом взаимодействии. В то же время вышла работа Ли, Р.~Оме и
Янга (США) \cite{47}, в которой также указывалось на то, что
корреляция спина и импульса (вскоре обнаруженная в эксперименте
мадам Ц.С.~Ву (США)) будет свидетельствовать как о нарушении
пространственной, так и о нарушении зарядовой четности.

Нарушение дискретных симметрий в слабом взаимодействии стало
причиной появления двух статей Ландау \cite{48}. Первая из них,
как в ней указывается, возникла из дискуссии с Л.~Окунем, Б.~Иоффе
и А.~Рудиком. В ней отмечено, что если предположить, что операция
P-инверсии должна сопровождаться заменой частиц на античастицы, то
законы природы остаются инвариантными относительно этого
преобразования, названного в статье комбинированной инверсией.
Подчеркивается, что в силу этой симметрии элементарные частицы не
могут иметь дипольных моментов. Как мы знаем сейчас,
комбинированная симметрия, или СР-четность, не является
фундаментальным законом природы: в 1964 году были обнаружены
распады долгоживущего нейтрального $K_L$-мезона на два
$\pi$-мезона, нарушающие СР. Тем не менее, само понятие
СР-симметрии оказалось исключительно плодотворным, как и поиски ее
нарушения в распадах $K$- и $B$-мезонов. Дипольные моменты
элементарных частиц в Стандартной Модели чрезвычайно малы, и на их
поиски до сих пор тратятся чрезвычайные усилия (см. ниже).

Во второй статье замечено, что нарушение Р-четности приводит к
возможности существования новых свойств у безмассового нейтрино. В
случае нулевой массы фермиона уравнение Дирака распадается на два
не связанных уравнения, переходящие друг в друга при инверсии
(уравнения Вейля). При отсутствии Р-инвариантности нейтрино может
описываться одним уравнением Вейля. Тогда нейтрино будет всегда
продольно поляризованным, а антинейтрино будет поляризовано
противоположным образом. Из экспериментальных данных по спектру
электронов, образующихся в распаде мюона, делается вывод о том,
что в этом распаде рождается пара $\nu \bar\nu$. Из операторов
продольных нейтрино и антинейтрино можно составить только
четырехмерный вектор, при этом из операторов мюона и электрона
можно составить две комбинации -- вектор и псевдовектор. Найдено
распределение вылетающих электронов по энергии и углу между
направлениями движения электрона и мюона (последнее направление
задает поляризацию мюона, рожденного в $\pi \to \mu\nu$-распаде).
Отметим, что при построении Стандартной Модели все фермионные
волновые функции выбираются в виде вейлевских спиноров. Массы
первоначально безмассовым фермионам дает механизм Хиггса. Теория
двухкомпонентного нейтрино одновременно была предложена Ли и Янгом
и А.~Саламом (Великобритания).

В 1957 году Л.Б.~Окунь и Б.М.~Понтекорво (ОИЯИ) заметили, что
малая разность масс $K_1^0$ и $K_2^0$ означает отсутствие
переходов с $\Delta S = 2$ в первом порядке по слабому
взаимодействию \cite{50}. В 1960 году Л.Б.~Окунь отметил, что
вклад второго порядка определяется величиной обрезания $\Lambda$,
которая должна быть порядка одного ГэВ'а \cite{51}. Надежду на то,
что столь низкое обрезание будет обеспечено сильным
взаимодействием, не удалось реализовать: в работе Б.Л.~Иоффе и
Е.П.~Шабалина 1967 года \cite{52} показано, что сильные
взаимодействия не приводят к обрезанию амплитуд процессов, идущих
во втором порядке по слабому взаимодействию. Как мы знаем сейчас,
это обрезание обеспечивается относительно малой массой $c$-кварка,
$m_c \approx 1.3$ ГэВ, а соответствующий механизм GIM был
предложен в 1970 году Ш.Л.~Глэшоу, Дж,~Илиопулосом и Л.~Майани
(США, Франция, Италия) еще до открытия $c$-кварка. 

После открытия
$b$-кварка в работе 1980 года М.И.~Высоцкий вычислил амплитуду
$K^0 - \bar K^0$-перехода в модели шести кварков, не предполагая
малости массы $t$-кварка по сравнению с массой $W$-бозона
\cite{V} (в литературе описывающие эту амплитуду функции
получили название функций Инами-Лима \cite{531,532}).
В главном логарифмическом приближении были найдены
также глюонные поправки к этой амплитуде. Отсутствие декаплинга
(``отцепления'') тяжелых частиц в электрослабой теории ярко
демонстрирует амплитуда $K^0 - \bar K^0$-перехода, в которой
имеются растущие при $m_t \gg M_W$ как $m_t^2$ вклады. Именно они
определяют нарушение СР в смешивании $K^0 - \bar K^0$ в рамках
шестикварковой модели Кобаяши--Маскава. Эти же вклады отвечают за
смешивание $B^0 - \bar B^0$. Открытие в 1986 году в DESY
коллаборацией АРГУС при активном участии экспериментаторов ИТЭФ
неожиданно большого $B^0 - \bar B^0$ смешивания дало первое
указание на аномально большую массу $t$-кварка.

В 1957--1958 гг. Л.Б. Окунь предложил составную модель, в которой
все известные в то время адроны (этот общепринятый сейчас термин
был введен в научный оборот Л.Б.~Окунем в 1962 г.) предлагалось
строить из трех ``прачастиц'' \cite{54}. В этом было отличие от
более ранней модели Сакаты, в которой адроны были построены из
физических частиц -- протона, нейтрона и лямбда-гиперона. На
основании этой модели предсказано существование нонета
псевдоскалярных мезонов и свойства двух его недостающих частиц
($\eta$- и $\eta^\prime$-мезонов). Сформулированы правила отбора
для полулептонных распадов странных частиц: $|\Delta S| = 1$,
$\Delta Q = \Delta S$, $\Delta T = 1/2$. На основе SU(3)-симметрии
сильного взаимодействия в 1962 году И.Ю.~Кобзарев и Л.Б.~Окунь
получили следствия для лептонных распадов мезонов \cite{55}.
Полный анализ лептонных распадов мезонов и барионов был проделан
Н.~Кабиббо (Италия) в 1963 г. Модель легла в основу известной
монографии ``Слабое взаимодействие элементарных частиц''
(Л.Б.~Окунь, первое издание 1963 г.) \cite{56}. Эта модель была
непосредственной предшественницей модели кварков.

Современная калибровочная теория электрослабых взаимодействий
изложена в монографии Л.Б.~Окуня ``Лептоны и кварки`` \cite{O}.  

В работе М.В.~Терентьева \cite{T} доказано отсутствие линейных по
нарушению изотопической симметрии поправок к слабому векторному току.
Это утверждение позволяет определить численное значение элемента 
матрицы Кобаяши-Маскава $V_{ ud}$ из анализа $\beta$-распадов ядер,   
обусловленных векторным током. В литературе оно известно как
теорема Адемолло-Гатто - аналогичное утверждение относительно 
нарушения SU(3) симметрии, доказанное для меняющего странность
векторного тока \cite{AG}.

В работе 1974 года М.Б.~Волошина, И.Ю.~Кобзарева и Л.Б.~Окуня был
впервые рассмотрен в рамках квантовой теории поля вопрос о распаде
ложного вакуума \cite{57}. Его актуальность связана с тем, что
экстраполяция хиггсовского потенциала Стандартной Модели в область
планковских значений поля Хиггса при современных значениях масс
$t$-кварка и бозона Хиггса приводит к утверждению о
метастабильности вакуума электрослабой теории.

А.И.~Вайнштейн (ИЯФ, Новосибирск), В.И.~Захаров и М.А.~Шифман
построили в 1978 году эффективный гамильтониан нелептонных слабых
распадов с учетом обменами глюонами в главном логарифмическом
приближении \cite{58}. Был обнаружен новый механизм усиления
переходов с $\Delta T = 1/2$, обусловленный так называемыми
``пингвинными'' диаграммами -- $s\to d$ переход с изучением
глюонов. Эти результаты в дальнейшем были обобщены на случай шести
кварков и имеют многочисленные приложения.

При исследовании моделей техницвета М.Б.~Волошин, Л.~Заскинд
(США), В.И.~Захаров, П.~Сикиви (США) обнаружили глобальную SU(2)-
симметрию в модели Глэшоу--Вайнберга--Салама, которая остается
ненарушенной при выпадении конденсата поля Хиггса \cite{59}. Эта
симметрия, названная в литературе ``custodual'' (``охранная''),
отвечает за близость масс $W$- и $Z$-бозонов. Ее важность стала
ясна при изучении радиационных поправок в электрослабой теории:
именно нарушение охранной симметрии большой массой $t$-кварка
позволило извлечь ее значение из прецизионных данных о параметрах
$W$- и $Z$-бозонов с точностью $\pm 30$ ГэВ, что облегчило
экспериментальное открытие $t$-кварка на Тэватроне ФНАЛ (США) в
1995 году.

Е.П.~Шабалин в 1978 году показал, что бытовавшая в литературе
оценка величины дипольного момента нейтрона в шестикварковой
модели CP-нарушения Кобаяши--Маскава ошибочна: суммарный эффект
двухпетлевых диаграмм равен нулю (зануление $d_n$ в одной петле
очевидно) \cite{60}. Ненулевой дипольный момент возникает на
уровне трех петель, и, согласно вычислению И.Б.~Хрипловича (ИЯФ,
Новосибирск), он на много порядков меньше современного
экспериментального ограничения \cite{61}. Этот факт служит сильным
стимулом к поиску ненулевого $d_n$: его обнаружение будет
свидетельствовать о существовании Новой Физики за рамками
Стандартной Модели.

В работе 1980 года И.Ю.~Кобзарев, Л.Б.~Окунь, Б.В.~Мартемьянов и
М.Г.~Щепкин параметризовали матрицу смешивания нейтрино в наиболее
общем случае наличия как дираковских, так и майорановских массовых
членов \cite{62}.

Обсуждавшаяся в литературе антикорреляция потока солнечных
нейтрино и активности Солнца в работах М.Б.~Волошина,
М.И.~Высоцкого и Л.Б.~Окуня \cite{63,64} была связана с возможным
проявлением магнитного момента нейтрино. Эти работы стимулировали
постановку ряда экспериментов по поиску магнитного момента
нейтрино в России и за рубежом.

В работе 1978 года Б.Л.~Иоффе и В.А.~Хозе (ПИЯФ) предложили искать
бозон Хиггса в реакции $e^+ e^- \to ZH$ \cite{65}, что и было
осуществлено на ускорителе LEP в ЦЕРН'е. Отрицательный результат
поисков привел к нижнему ограничению на массу: $m_H > 114$ ГэВ.

Высокая точность измерения параметров $Z$-бозона на $e^+
e^-$-коллайдерах LEPI (ЦЕРН) и SLC (СЛАК) и массы $W$-бозона на
$e^+ e^-$-коллайдере LEP II и Тэватроне позволила произвести
проверку электрослабой теории с учетом радиационных поправок.
Перенормируемость электрослабой теории делает возможным вычисление
таких поправок. М.И.~Высоцким, В.А.~Новиковым, Л.Б.~Окунем и
А.Н.~Розановым были получены соответствующие формулы, выражающие
измеряемые на эксперименте параметры через наиболее точно
измеренные значения фермиевской константы $G_F$, массу $Z$-бозона
$M_Z$ и постоянную тонкой структуры на масштабе массы $Z$-бозона
$\alpha(M_Z)$ \cite{66,67,671}. На первом этапе (1991--1995 гг.)
полученные результаты использовались для предсказания массы
$t$-кварка. После открытия $t$-кварка на Тэватроне и измерения его
массы было получено предсказание для массы бозона Хиггса в рамках
Стандартной Модели $M_H = 80^{+30}_{-20}$ ГэВ, подтвердившееся ее
измерением в 2012 году на LHC (ЦЕРН): $M_H = 125 \pm 1$ ГэВ.


\section*{Сильные взаимодействия}

Научная слава теоротдела во многом обязана фундаментальным
результатам Померанчука, полученным в этой области. В 1958 году,
исходя из дисперсионных соотношений, им доказано асимптотическое
равенство полных сечений взаимодействий частиц и античастиц с
фиксированной мишенью (теорема Померанчука) \cite{68}. В работах
В.Н.~Грибова (ПИЯФ) и Померанчука квантовомеханическая теория
полюсов Редже была использована для создания последовательной
картины процессов при асимптотически высоких энергиях \cite{69}. В
честь Померанчука реджевский полюс с квантовыми числами вакуума,
ответственный за выполнение теоремы Померанчука, получил за
рубежом название ``померон''. Дальнейшее развитие реджистика
получила в работах К.А.~Тер-Мартиросяна и А.Б.~Кайдалова.

После создания КХД и открытия асимптотической свободы в сильных
взаимодействиях теоретиками ИТЭФ был получен ряд фундаментальных
результатов, ставших классическими. Так, в работе \cite{NShVZ}
была предложена наивная кварковая модель для глубоко-неупругих
процессов, позволяющая ``сшить'' составную модель нуклонов с
распределениями кварков и глюонов при больших переданных
импульсах. В дальнейшем эти идеи были развиты в работах
европейских теоретиков.

В работе \cite{NShVZ2}  было осознано, что масса $c$-кварка $m_c$
велика в масштабах $\Lambda_{\small\rm QCD}$ и можно учесть
сильные взаимодействия в процессах с тяжелыми кварками,
раскладывая по малой константе связи $\alpha_s(m_c)$. Так были
вычислены сильные поправки к разности масс $K_L$ - $K_S$ к фото- и
электророждению чарма.

После открытия $J/\psi$-мезона в работах А.И.~Вайнштейна,
М.Б.~Волошина, В.И.~Захарова, В.А.~Новикова, Л.Б.~Окуня и
М.А.~Шифмана была построена дисперсионная теория чармония и
написаны знаменитые обзоры \cite{70,701}, ставшие настольной книгой для
всех, кто занимался физикой тяжелых кварков. Дальнейшее развитие
этих идей А.И.~Вайнштейном, В.И.~Захаровым и М.А.~Шифманом
\cite{71} привело к знаменитым правилам сумм КХД, позволившим
вычислять свойства адронов, построенных из легких $u$-, $d$- и
$s$-кварков, в терминах вакуумных конденсатов. В работах
В.М.~Беляева, Б.Л.~Иоффе, Я.И.~Когана, В.Л.~Елецкого и А.В.~Смилги
этот подход был использован в различных приложениях \cite{72}.

В работе 1976 года М.Б.~Волошин и Л.Б.~Окунь обсуждали возможность
существования кварковых ``молекул'' -- связанных состояний
мезонов, содержащих тяжелые $c$-кварки \cite{73}. В последние годы
такие состояния были обнаружены в системах $c$- и $b$-кварков. Их
свойства активно обсуждаются в литературе; М.Б.~Волошиным
развивается ``молекулярный'' подход; имеются и другие подходы к
проблеме (в ИТЭФ физикой экзотических адронов занимаются
А.М.~Бадалян, Ю.С.~Калашникова, А.Е.~Кудрявцев, А.В.~Нефедьев и
Ю.А.~Симонов \cite{74}).

\section*{Точные результаты в КТП}

Теоретиками ИТЭФ было получено несколько точных результатов в
квантовой теории поля. Так, в работе Е.Б. Богомольного \cite{Bog}
был найден новый класс решения классических уравнений для теории
калибровочных полей со скалярным конденсатом. Это так называемые
BPS-монополи (Богомольный--Прасад--Зоммерфельд). Эти решения
играют исключительно важную роль в $N=2$ и $N=4$ суперсимметричных
теориях. В работах \cite{NShVZ3} были получены многочисленные
точные соотношения для корреляторов и показано, что в сильных
взаимодействиях глюболов и гибридных состояний масштаб масс может
на порядки превышать $\Lambda_{\small\rm QCD}$.

В работах \cite{NShVZ4} для суперсимметричных теорий был построен
формализм для описания суперинстантонов. Были доказаны теоремы об
отсутствии поправок к инстантонным амплитудам. В результате был
точно вычислен глюинный конденсат для суперсимметричных КХД и
точная $\beta$-функция (так называемая NSVZ $\beta$-функция). В
дальнейшем принстонской группой было показано, что суперинстантоны
могут приводить к динамическому нарушению суперсимметрии.

\section*{Гравитация, космология}

В работе В.В.~Судакова, Е.М.~Лифшица (ИФП) и И.М.~Халатникова
(ИФП) 1961 года обсуждалась
сингулярность в основанных на ОТО космологических моделях
\cite{75}.

В работе В.И.~Захарова 1970 года
\cite{76} было показано, что введение сколь угодно малой массы
гравитона противоречит наблюдательным следствиям ОТО (т.н.
сингулярность Вельтмана--Ван Дама--Захарова). Механизм построения теории
гравитации, которая имеет непрерывный предел к нулевой массе гравитона,
был предложен в работе А.И.~Вайнштейна \cite{77} и обсуждался в
целом ряде последующих работ.

В работе Зельдовича, Окуня и Пикельнера~\cite{ZOP} в 1965 году была вычислена космологическая концентрация
реликтовых кварков для случая, если бы они могли существовать в свободном виде. Полученный результат
противоречил имевшимся ограничениям на обилия кварков во вселенной и таким образом однозначно
свидетельствовал в пользу невылетания (конфайнмента) кварков. В этой работе было использовано
кинетическое уравнение для расчета космологической концентрации тяжелых частиц, которое было
впоследствии, в 1977 году, переоткрыто Б. Ли и С. Вейнбергом и носит их имя.

Кобзаревым, Окунем и Померанчуком была предложена и рассмотрена идея о зеркальной материи~\cite{KOP-mirror}.
Это была пионерская теоретическая работа о возможном существовании темного вещества во вселенной.
Впоследствии идеи о зeркальной темной материи развивались в ИТЭФ С.И. Блинниковым,
см. обзор~\cite{SIB-mirror}.

В работах Зельдовича, Кобзарева и Окуня~\cite{ZKO-CP} было обнаружено, что модель спонтанного нарушения СР-симметрии
прoтиворечит наблюдаемым данным об изотропии вселенной, т.к. гигантская плотность энергии стенки разделяющей
домены вещества и антивещества разрушит изотропию микроволнового фона. Эта работа инициировала исследования
возможных механизмов "рассасывания" доменных стенок. Возможные решения этой проблемы описаны в недавней
работе~\cite{DRT-walls}.

Вычисление космологической концентрации слабо взаимодействующих реликтовых частиц
во Вселенной было проведено М.И.~Высоцким, А.Д.~Долговым и Я.Б.~Зельдовичем
\cite{78}  в 1977 году. Практически одновременно появились аналогичные работы
Б. Ли и С.~Вайнберга (США), Хута (Нидерланды) и К.~Сато и М. Кобаяши (Япония)).
Результаты этих работ являются
основополагающими при расчетах плотности массивных частиц темной материи, в частности,
нейтралино.

А.Д. Долговым и Я.Б. Зельдовичем в 1980 году был написан широко известный обзор
"Космология и элементарные частицы" \cite{AD-YBZ-rev},
содержащий значительное количество оригинальных результатов. Этот обзор в заметной степени инициировал
развитие этой области, в частности, на Рочестерких конференциях появилась секция, повторяющая название
обзора.

Значительный вклад в исследование проявлений нейтрино в космологии был внесен А.Д. Долговым с соавторами.
Было выведено кинетическое уравнение для матрицы плотности осциллирующих нейтрино~\cite{AD-density-matr},
которое в настоящее время является основным инструментом при изучении эффектов нейтринных осцилляций в ранней
вселенной и при взрыве сверхновых. На основе решения этого уравнения Барбиери и Долговым~\cite{RB-AD}
были получены пионерские ограничения на пaраметры осцилляций нейтрино по наблюдаемым обилиям легких
элементов. Этот результат был позднее обобщен на более реалиситческий случай перемешивания всех
активных нейтрино~\cite{AD-FV}. Кроме того, в цитированных выше работах~\cite{RB-AD},
см. также~\cite{AD-nu-rev},  был разработан метод вычисления космологичeской
концентрации  стерильных нейтрино, который лежит в основе вычислений плотности теплой темной материи, если
она состоит из стерильных нейтрино. Этот метод получил название метода Додельсона-Видроу по более поздней работе
указанных авторов. В работах~\cite{AD-k-nu} был обнаружен новый эффект разогрева нейтрино за счет поздней аннигиляции
более горячих электронно-позитронных пар. Согласно результатам этих работ стало общепринятым, что каноническое
эффективное число типов нейтрино в космологии не три, как можно бы наивно ожидать, а 3.046. Долговым с соавторами
было получено наиболее сильное ограничение на химический потенциал космологических нейтрино~\cite{AD-chem-pot},
которое исключает влияние нейтринного вырождения на структуру вселенной и СМВ, как считалось до того.

А.Д. Долговым с соавторами проведены пионерские вычисления по разогреву вселенной после инфляции, как в рамках
теории возмущений~\cite{AD-AL}, так и непертубативно~\cite{AD-DK}.

В работе Блинникова с соавторами~\cite{SIB-Hubble} был предложен новый метод прямого определения параметра
Хаббла, основанный на наблюдении сверхновых типа IIn. Этот метод свободен от неоднозначностей классического
подхода к построению лестницы космических расстояний. Его применение к известным сверхновым, расстояние до
которых было надежно измерено другими способами, показало замечательное согласие нового подхода с
традиционными.

В цикле работ Долгова с соавторами изучалась теория модифицированной F(R) гравитации, предложенной в
качестве одого из возможных механизмов объяснения ускоренного расширения вселенной. Был открыт эффект
сильной неустойчивости большого класса таких теорий~\cite{AD-MK}, получивший в литературе название
неустойчивости Долгова-Кавасаки. В результате потребовалось изменение  модифицированных
теорий, чтобы избежать этой неустойчивости. При дальнейших исследованиях таких теорий был открыт целый
ряд новых явлений, таких как: высокочастотные осцилляции скаляра кривизны $R$~\cite{EA-AD}, антигравитация
(гравитационное отталкивание) в системах конечного размера~\cite{EA-AD-LR-anti}, новые эффекты в гравитационной
неустойчивости, отличающиеся от известной неустойчивости Джинса~\cite{EA-AD-LR-Jeans} .
Проверка этих предсказаний позволит либо однозначно подтвердить модель, либо получить ограничения на
ее параметры, либо даже отвергнуть ее.

\section*{Кинетика и термодинамика}

После открытия в 1964 году нарушения СР в распадах К-мезонов стало практически очевидным, что
в силу СРТ-теоремы также нарушается T-инвариантность, т.е. инвариатность относительно обращения
времени. Тогда возник вопрос о справедливости канонических равновесных распределений в квантовой
статистике, которые стандартным образом выводились на основе условия детального баланса,
следующего из Т-инвариантности. В работе А.Д. Долгова \cite{AD-kin-equil} в этой связи
было показано, что "равновесная кинетика сильнее T-нарушения". Унитарность S-матрицы обеспечивает
справедливость стандартных равновесных распределений, хотя условие детального баланса и нарушается.
Равновесие между прямыми и обратными реакциями достигается за счет нескольких цепочек реакций. По
этой причине на смену детальному балансу при T-нарушении приходит циклическй баланс, как названо в
цитируемой выше работе.

М.И.Высоцкий и В.А.Новиков поддержаны грантами РФФИ 14-02-00995, 16-02-00342 и 
НШ-6792.2016.2, А.Д.Долгов -
грантом правительства РФ 11.Г34.31.0047.


\end{document}